\documentclass[journal=jacsat,manuscript=article]{achemso}
\setkeys{acs}{usetitle = true}
\usepackage{siunitx} 
\usepackage[version=3]{mhchem} 
\usepackage[T1]{fontenc}       
\usepackage{graphicx}
\usepackage{dcolumn}
\usepackage{threeparttable}
\usepackage[usenames, dvipsnames]{color}
\usepackage{gensymb}
\usepackage{inputenc}  
\usepackage{physics}
\usepackage{bm} 
\usepackage{amsmath}  
\usepackage{amsfonts} 

\title{Deterministic coupling of quantum emitters in WSe$_{2}$ monolayers to plasmonic nanocavities}

\author{Oliver Iff}
\affiliation[]{Technische Physik and Wilhelm Conrad R\"ontgen Research Center for Complex Material Systems, Physikalisches Institut,
Universit\"at W\"urzburg, Am Hubland, D-97074 W\"urzburg, Germany}
\author{Nils Lundt}
\affiliation[]{Technische Physik and Wilhelm Conrad R\"ontgen Research Center for Complex Material Systems, Physikalisches Institut,
Universit\"at W\"urzburg, Am Hubland, D-97074 W\"urzburg, Germany}
\author{Simon Betzold}
\affiliation[]{Technische Physik and Wilhelm Conrad R\"ontgen Research Center for Complex Material Systems, Physikalisches Institut,
Universit\"at W\"urzburg, Am Hubland, D-97074 W\"urzburg, Germany}
\author{Laxmi Narayan Tripathi}
\affiliation[]{Department of Physics, Birla Institute of Technology, Mesra, Ranchi 835215, Jharkhand, India}
\alsoaffiliation[Second University]{Technische Physik and Wilhelm Conrad R\"ontgen Research Center for Complex Material Systems, Physikalisches Institut,
Universit\"at W\"urzburg, Am Hubland, D-97074 W\"urzburg, Germany}
\author{Monika Emmerling}
\affiliation[]{Technische Physik and Wilhelm Conrad R\"ontgen Research Center for Complex Material Systems, Physikalisches Institut,
Universit\"at W\"urzburg, Am Hubland, D-97074 W\"urzburg, Germany}
\author{Young Jin Lee}
\affiliation[]{Department of physics, Chung-Ang University,  Seoul, South Korea}
\author{Soon-Hong Kwon}
\affiliation[]{Department of physics, Chung-Ang University,  Seoul, South Korea}
\author{Sven H\"{o}fling}
\affiliation[]{Technische Physik and Wilhelm Conrad R\"ontgen Research Center for Complex Material Systems, Physikalisches Institut,
Universit\"at W\"urzburg, Am Hubland, D-97074 W\"urzburg, Germany}
\alsoaffiliation[Second University]
{SUPA, School of Physics and Astronomy, University of St. Andrews, St. Andrews KY16 9SS, United Kingdom}
\author{Christian Schneider}
\email{christian.schneider@physik.uni-wuerzburg.de}
\affiliation[]{Technische Physik and Wilhelm Conrad R\"ontgen Research Center for Complex Material Systems, Physikalisches Institut,
Universit\"at W\"urzburg, Am Hubland, D-97074 W\"urzburg, Germany}

\begin{document}
\begin{abstract}
We discuss coupling of site-selectively induced quantum emitters in exfoliated monolayers of WSe$_2$ to plasmonic nanostructures. Squared and rectangular gold nanopillars, which are arranged in pitches of \SI{4}{\micro\meter} on the surface, have sizes of tens of nanometers, and act as seeds for the formation of quantum emitters in the atomically thin materials. 
We observe chraracteristic narrow-band emission signals from the monolayers, which correspond well with the positions of the metallic nanopillars with and without thin dielectric coating. Single photon emission from the emitters is confirmed by autocorrelation measurements, yielding $g^{2}(\tau=0)$ values as low as 0.17. Moreover, we observe a strong co-polarization of our single photon emitters with the frequency matched plasmonic resonances, indicating deterministic light-matter coupling. Our work represents a significant step towards the scalable implementation of coupled quantum emitter-resonator systems for highly integrated quantum photonic and plasmonics applications. 
\end{abstract}

\section{Introduction}

Creating solid state quantum emitters and their integration in micro- and nanophotonic structures is one of the prime tasks in modern quantum engineering. Coupled solid state quantum emitter-cavity systems range among the most promising candidates for the realization of highly efficient single photon sources \cite{Ding2016,Schlehahn2016,Unsleber2015,Senellart2017,Chang2006}, spin photon interfaces \cite{DeGreve2012,Gao2012}, quantum sensing probes \cite{Anker2008} as well as building blocks for quantum simulation \cite{Wang2017d} and surface code quantum computing \cite{Greve2013}. While quantum emitters have been identified, studied and engineered in a variety of crystals including III-V \cite{Michler2000} and II-VI quantum dots \cite{Peng2000,Lowisch1996,Xin1996}, colour defects in diamonds \cite{Kurtsiefer2000}, impurities in SiC and organic polymers \cite{Castelletto2013,Castelletto2014}, atomically thin materials \cite{He2015,Srivastava2015a,Tonndorf2015,Koperski2015, Kumar2015} were recently established as a novel platform of quantum photonic devices.
Quantum dots in III-V semiconductors and defect centers in diamonds certainly belong to the most mature implementations \cite{Balasubramanian2008}, but the quality of site-controlled emitters leaves still needs to be improved, putting a serious thread regarding their scalable fabrication in ordered arrays \cite{VanDerSar2009}. Ordered InAs/GaAs quantum dot arrays have been realized by selective area growth methods and epitaxial growth on patterned substrates \cite{schmidt2007lateral}, but in most cases the costly fabrication methods severly compromised their emission properties. Direct integration of positioned solid state quantum emitters with photonic resonators has been accomplished \cite{Gallo2008,Schneider2009,Sunner2008a}, but only in very selected cases and genuine scalability has remained elusive. 

The formation of quantum emitters in mono- and bilayers of transition metal dichalcogenides has now been observed in various implementations: initially, localized luminescence centers in exfoliated flakes were discovered close to their edges, and have been associated with strain wrinkles \cite{Tonndorf2015,Koperski2015, Kumar2015}. In epitaxially grown flakes, random positioning of such spots was observed \cite{He2015}, indicating emission from defect bound excitons. Recently, the formation of quantum emitters on modulated metal substrates \cite{Kern2015,Tripathi2018}, as well as nanopillars \cite{Palacios-Berraquero2017,Branny2017} was reported and associated with localized and engineered crystal strain fields, which outlines the unique possibility to deterministically induce quantum emitters in a straight forward manner by structuring the sample surface prior to the transfer. 

While the ordered formation of quantum emitters thus far has been mainly observed on dielectric, nanostructured surfaces, spontaneous emission enhancement was reported on rough metallic surfaces and gold-coated nanopillars, giving rise to localized plasmonic modes \cite{Tripathi2018,Cai2018}. Combining atomically thin materials which comprise either tightly localized excitons or strongly bound free excitons with nanoplasmonic cavities yields a promising pathway to study light-matter coupling on the nanoscale enabled by the enormous field enhancements provided by metallic nanostructures \cite{Krasnok2018, Lalanne2018}. However, the deterministic coupling of well-ordered quantum emitters in atomically thin materials with resonant plasmonic modes has only now been achieved \cite{Luo2018,Cai2018}. 

\section{Sample Structure and Setup}
In this work, we demonstrate the feasibility to induce ordered arrays of quantum emitters by defined arrays of metallic nanopillars, fabricated on a SiO$_2$ substrate. Such structures directly represent a coupled quantum dot-nanocavity system, and act as polarization-controlled single photon sources.  

\begin{figure}
	\centering
	\includegraphics[scale=0.6]{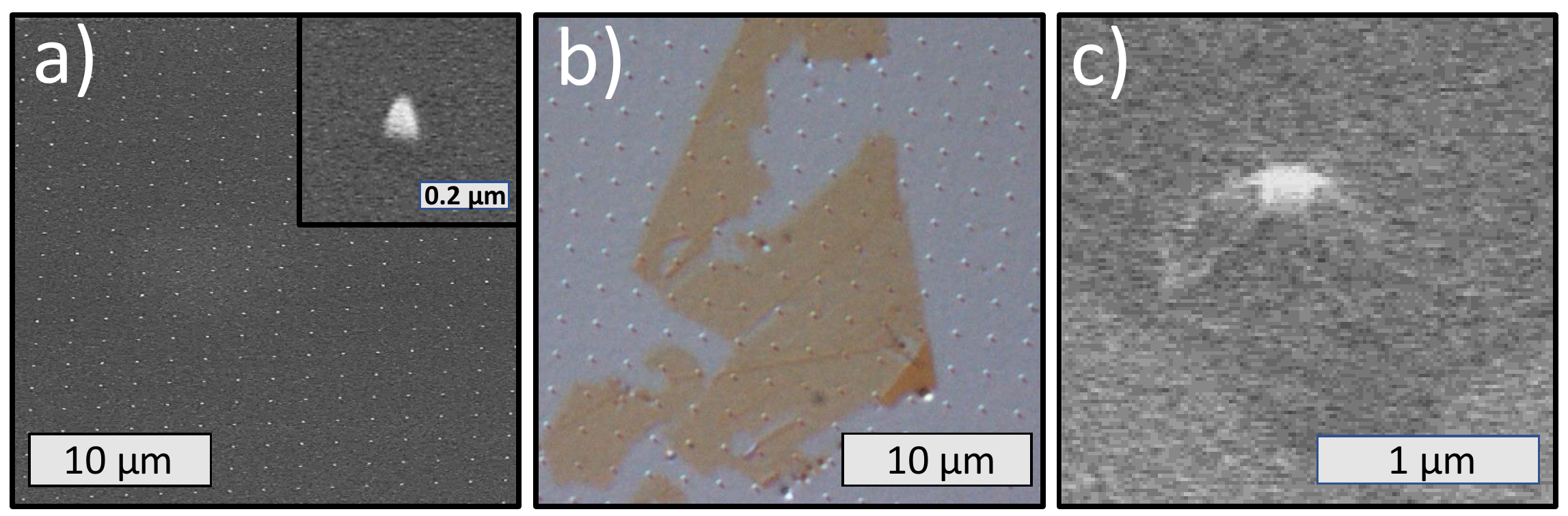}
	\caption[]{\label{Surface structure and modulated monolayer}(a) Scanning electron microscope (SEM) image of the sample surface comprising metallic nanopillars as quantum emitter seeds and plasmonic nano-cavities. Inset: close-up view of a nanopillar. (b) Optical image of the pillar array after successfull dry-transfer of an atomically thin WSe$_{2}$ monolayer. (c) Close-up SEM image of a single pillar covered by a strained monolayer, showing the formation of wrinkles.}
\end{figure}

The sample consists of a semi-insulating silicon substrate, with a \SI{200}{\nano\meter} thick SiO$_2$ layer on top. In order to fabricate the nanopillars, we first spin-coated a thin layer of PMMA and performed electron beam lithography to selectively expose rectangular areas in the resist with dimensions of \SI{20}{\nano\meter} - \SI{240}{\nano\meter}. After developing the resist, a \SI{80}{\nano\meter}  thick gold layer was evaporated on the sample, followed by a lift-off step. A scanning electron microscope (SEM) image of a prototype nanopillar array with a pitch of \SI{2}{\micro\meter} is shown in Fig. 1a). On selected samples we additionally deposited a \SI{3}{\nano\meter} thin layer of Al$_2$O$_3$ via atomic layer deposition. 
Next, we fabricated atomically thin layers of WSe$_2$ via mechanical exfoliation using adhesive tape, and transferred the layers on the pillar arrays via dry transfer \cite{Castellanos-Gomez2014} (Fig. 1b). We observe, that part of the pillars pierced the monolayer, while a substantial number of nanopillars (> 50 $\%$) locally strained the layer, yielding the tent-like structure shown in Fig. 1c). 

Spatially resolved optical spectroscopy was performed in a micro-photoluminescence setup with optional high spatial resolution (using fiber based confocal setting). The sample is excited by a frequency-doubled Nd:YAG laser at \SI{532}{\nano\meter}, mounted in a liquid helium cooled flow cryostat. 

\section{Experimental Results and Discussion}

Fig. 2a) depicts an exemplaric power dependent luminescence spectrum recorded on the position of a nanopillar with dielectric coating. The spectrum is widely dominated by a zoo of sharp emission lines, a typical signature of strongly localized emission centers in the crystal. In the low-power regime, these emission lines exhibit a slightly sub-linear intensity increase with the pump power prior to their saturation level (Fig. 2b). This behaviour is mainly due to the gold pillars absorbing parts of the incident laser light \cite{Link1999}, but also the re-emitted light from the emitter, reducing their quantum efficiency.

 \begin{figure} 
	\centering
	\includegraphics[scale=0.7]{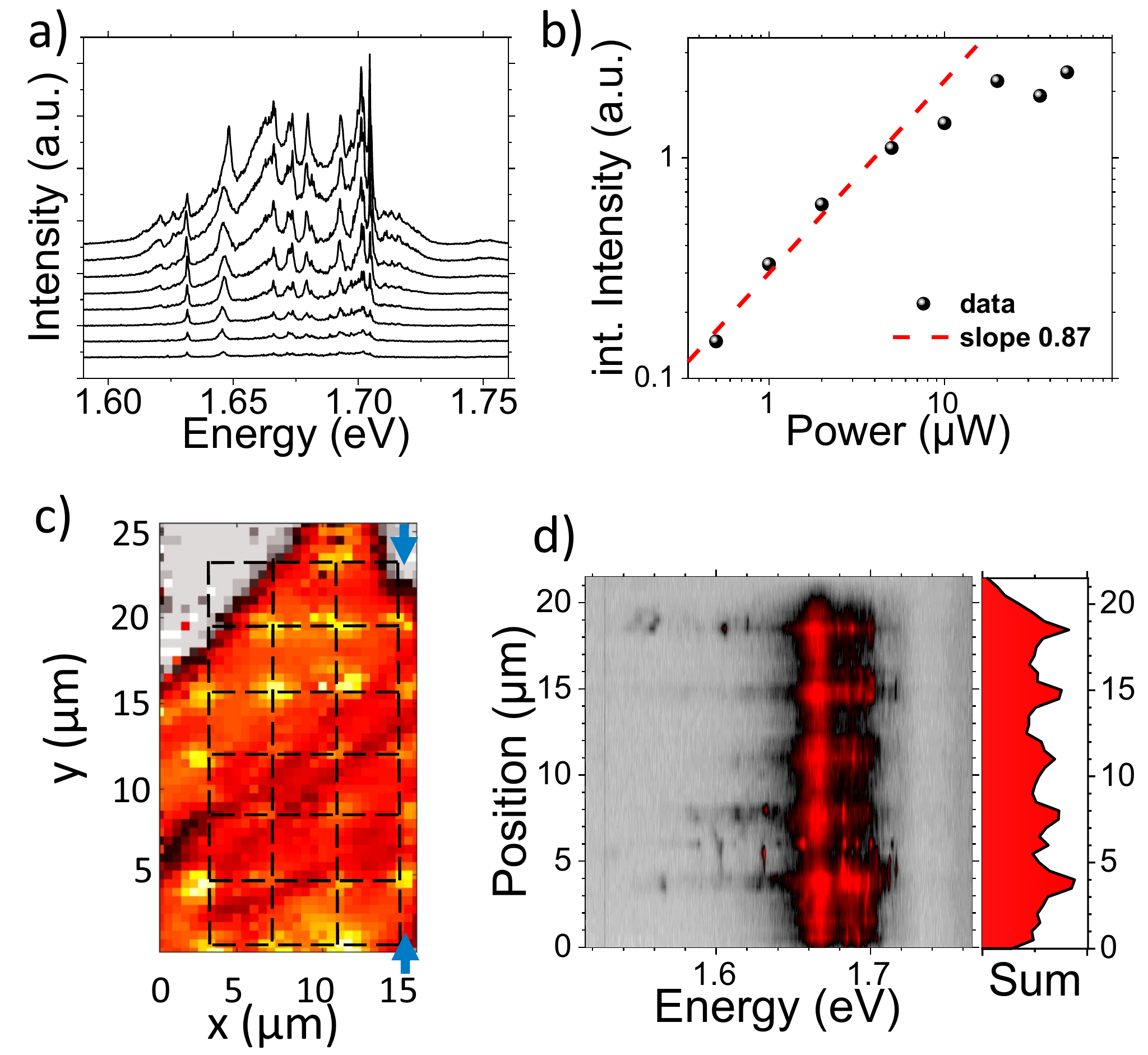}
	\caption[]{\label{Microphotoluminescence spectroscopy of localized excitons} a) Power-dependent spectra on a nanopillar revealing many discrete emitters. b) Power-dependent study of a quantum emitter emission line before saturation starts above \SI{10}{\micro\watt}. c) Spatial map of a WSe$_2$ flake covering the nano-pillar array, showing the integrated intensity from $700 - \SI{800}{\nano\meter}$. The enhanced PL coincides with the \SI{4}{\micro\meter} pillar distance (black pattern). d) Spectral information extracted between the blue arrows in c), revealing a periodic increase in luminescence and the formation of additional localized emission centers.}
\end{figure}

The ordered formation of emitters on the nanopillar arrays is confirmed in a highly spatially resolved scanning microphotoluminescence study, applying the confocal configuration: Here, we carefully scan the sample's surface by utilizing a pair of motorized linear stages with a step width of \SI{500}{\nano\meter} underneath the excitation and collection spot. The spectrally integrated map (700-\SI{800}{\nano\meter}) is shown in Fig. 2c). It  clearly evidences a regular pattern of bright emission sites, perfectly coinciding with the positions of the metallic pillars (dashed black pattern). Spectral information is best illustrated in a selected linescan between the blue arrows in Fig 2c). Here, we clearly observe a two-fold effect by the nanopillars (Fig. 2d): A strong luminescence enhancement of the overall signal due to plasmons \cite{Hugall2018}, as well as the regular formation of the sharp peaks below the free exciton energy (<\SI{1.74}{\electronvolt}), which we associate with tight exciton localization due to strain.\\


 \begin{figure} 
 	\centering
 	\includegraphics[scale=0.6]{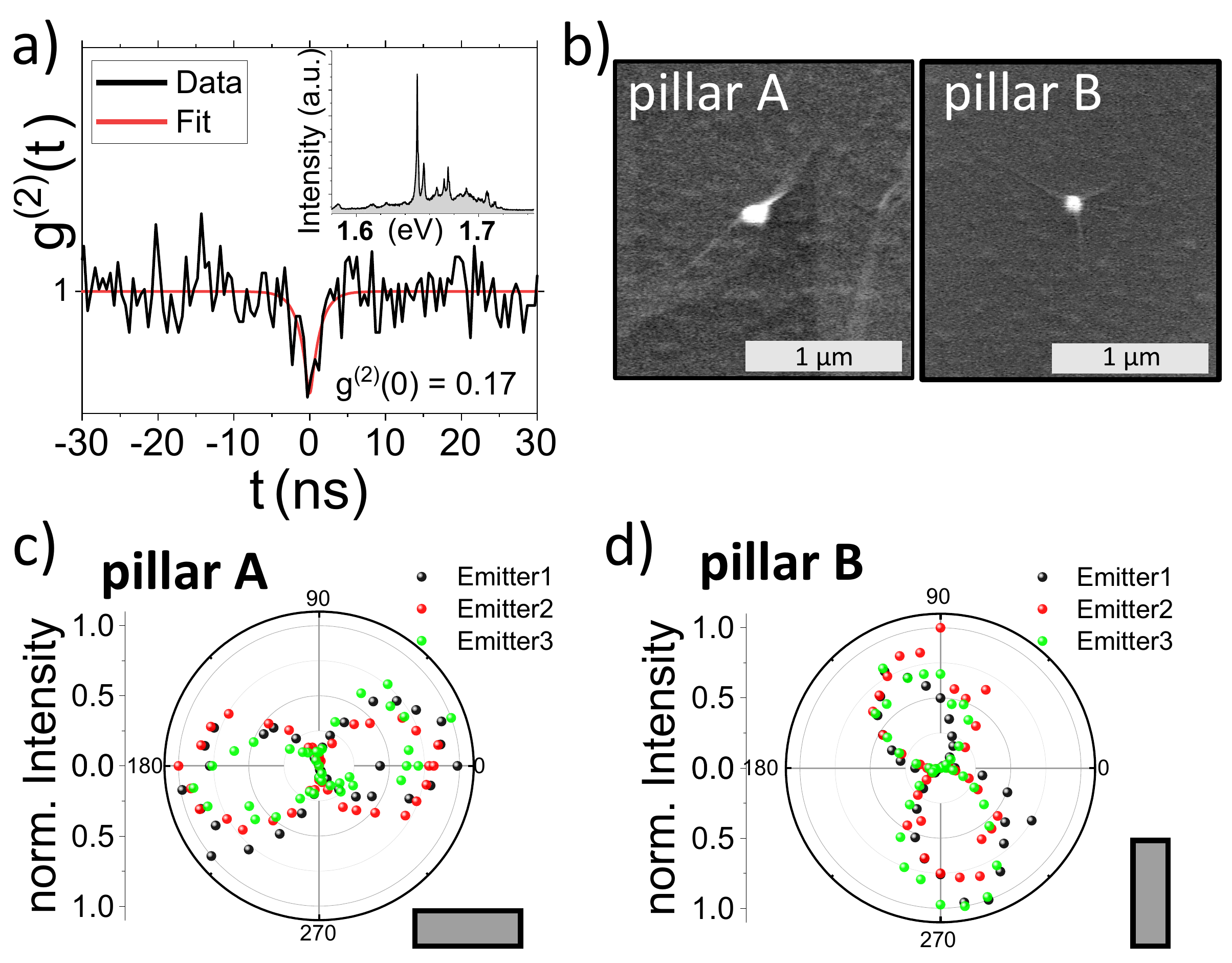}
 	\caption{\label{polarization and auto correlation} a) Second-order autocorrelation function of a quantum emitter on a pillar. The value of $g^{(2)}(\tau=0) = 0.17$ confirms single photon emission. Inset: spectrum of single photon emitter. b) SEM images of two individual rectangles covered by WSe$_2$. Pillar A is horizontally and pillar B vertically aligned. c) and d) Polarization characteristic of three individual quantum emitters each on two different \SI{90x30}{\nano\meter} nanopillars shown in b).}
 \end{figure}

In order to provide evidence for the capability to emit single photons from the deterministically localized excitons, we performed second order correlation  measurements by exciting the sample with a \SI{532}{\nano\meter} CW laser (Fig.~3a). We selected a dominant emission feature from one square pillar ($140 \times \SI{140}{\nano\meter}$). The luminesence was spectrally filtered (bandwidth: $\approx$ \SI{300}{\micro\electronvolt}, 300 grooves/mm grating) and passed to a fiber coupled Hanbury Brown and Twiss (HBT) setup. We observed a well-pronounced anti-bunching signal at zero delay time ($\tau = 0$), allowing us to extract a $g^{(2)}(\tau=0)$ value of 0.17, which clearly puts our system in the regime of single photon emission.
 
Polarization resolved spectroscopy on different nanopillars revealed a strongly linear polarization of the luminescence from the emitters. In Fig. 3b two exemplary \SI{90x30}{\nano\meter} pillars are shown which are aligned perpendicular to each other and covered by the monolayer. Comparing the polarization of several emitters from these two pillars shows a strong correspondence of the polarization and pillar orientation (Fig. 3c,d). This alignment of the polarization along the long axis of the given gold rectangle can be associated with the coupling of the emitter to the plasmonic excitations in the metal which are much more pronounced in the extended axes as has been demonstrated with similar plasmonic structures before \cite{Luo2017}. Slight modifications of the rotation angle also depends on the way the monolayer bends around the pillar, which further acts on the polarization of the emission \cite{Kern2015}.  

\begin{figure} 
	\centering
	\includegraphics[scale=0.6]{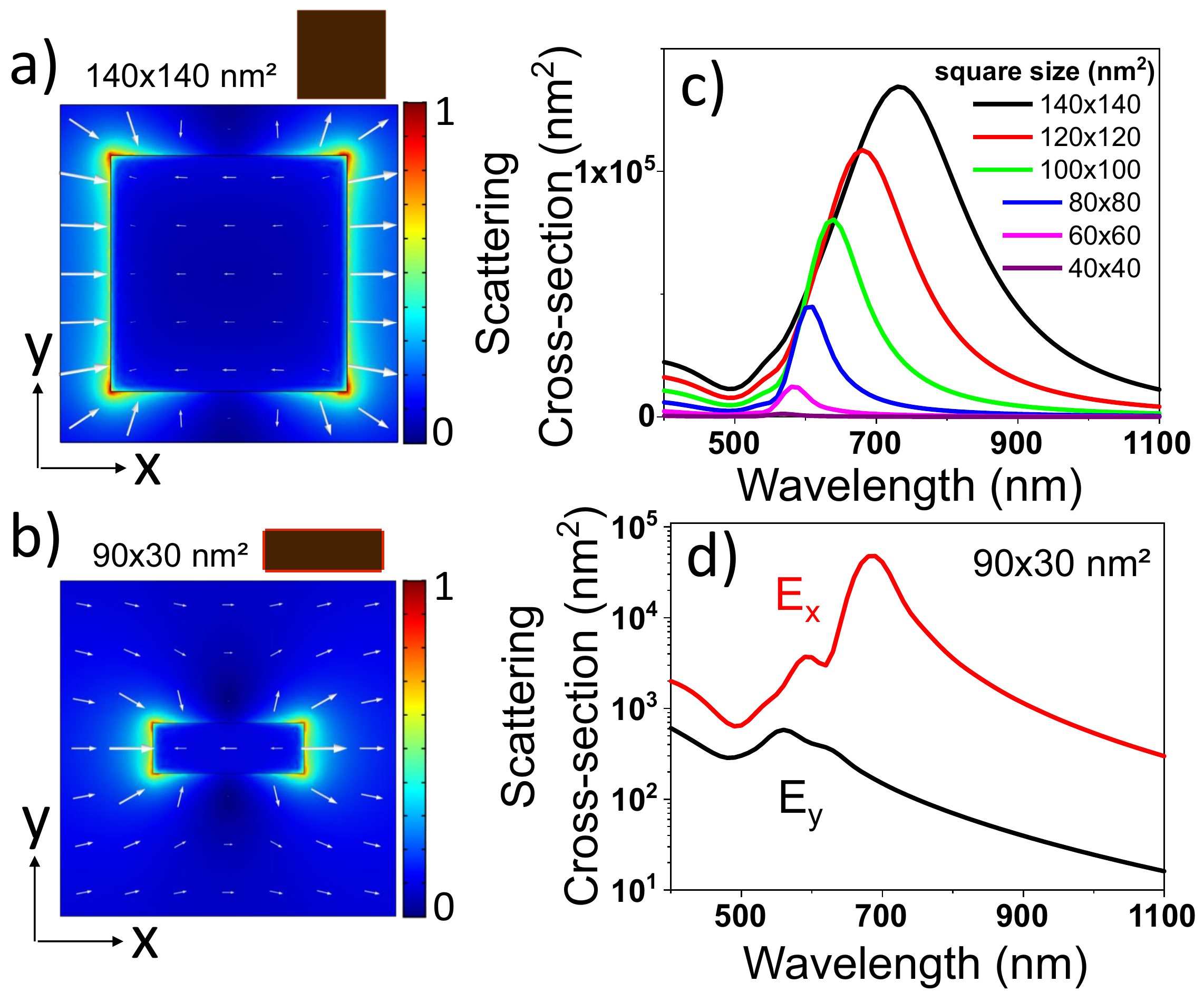}
	\caption{FDTD simulation: Vector maps of the field distribution and electric field enhancement E/E$_0$ at top surface of a pillar with a) a square cross-section of (\SI{140x140}{\nano\meter}) and a rod-like cross-section of (\SI{90x30}{\nano\meter}) excited by an electromagnetic field E$_0$ of 1 V/m at \SI{740}{\nano\meter}. c) Calculated scattering cross-section spectrum of square nanopillars with different size from \SI{140}{\nano\meter} to \SI{40}{\nano\meter}. d) Two scattering cross-section spectrums of a rod-like nanopillar with a size of \SI{90x30}{\nano\meter} for an incident light with orthogonal polarizations, E$_x$ (red) and E$_y$ (black). x-direction represents the direction of the long edge of the rod-like nanopillar.}
\end{figure}

\section{Theory}

In order to understand the optical enhancement of the quantum emitters in a WSe$_2$ via a metal nanopillar array, we investigated the plasmon modes excited in a square and a rod-like nanopillar by finite-difference time-domain (FDTD) method. In a square pillar with a size of \SI{140x140}{\nano\meter}, the electric field distribution on top surface of the pillar is calculated, as shown in Fig. 4(a), where a quantum emitter of WSe$_2$ can be placed. At two vertical side edges orthogonal to the E$_x$ polarization, strong field enhancement (E/E$_0$) with a maximum value of 20 compared with an incident light is observed. In addition, vector plot shows the field enhancement is strongly attributed from the E$_x$ field at the edges. Because an emitter in 2D materials oscillates in plane, strong emission enhancement for the defect emitters at the edges can be induced by the resonant coupling of the plasmon mode. The mode of the Fig. 4a is obtained by assuming E$_x$ linearly polarized incident light, and a 90-degree rotated mode (not shown in the Figure) can also be observed for E$_y$ polarized incident light.
Scattering cross-section spectrum of Fig. 4(c) represents the spectral dependence of the plasmonic mode in a square pillar for different sizes of the side edges from \SI{40}{\nano\meter} to \SI{140}{\nano\meter}, where the resonant peak for a size of \SI{140x140}{\nano\meter} is \SI{733}{\nano\meter} with a large FWHM of \SI{234}{\nano\meter}. Therefore, the radiative emission of the quantum emitter with a spectrum of Fig. 3(a) can be enhanced by resonant coupling with a plasmonic mode. The plasmon resonance shifts blue with decreasing the size.

In order to understand the strong polarization dependence of the emission from a rod-like nanopillar, the mode profile and the scattering cross-section of a rod-like nanopillar with a cross-section of \SI{90x30}{\nano\meter} are investigated. For the E$_x$ polarized incident light, the electric field distribution of the plasmonic mode in Fig. 4b is similar with that of a square pillar, Fig. 4a except for that the mode is elongated along the x-direction following the rod-like shape. In the scattering cross-section spectrum, the rod-like pillar exhibits strong plasmon resonance peak at a wavelength of \SI{686}{\nano\meter} for an E$_x$ polarized light, however, there are no significant resonances for an E$_y$ polarized light. Therefore, \SI{90x30}{\nano\meter} nanopillar can enhance the x-directionally polarized emission of a quantum emitter and suppress the y-directionally polarized emission, resulting in strong linear polarization aligned along the long axis, as shown in Figs. 3(c) and (d).

\section{Summary}

In conclusion, we demonstrated the formation of ordered arrays of quantum emitters in an atomically thin layer of WSe$_2$, transferred on a metal nanopillar array. The gold nanopillars yield the formation of quantum emitters, and furthermore can act as plasmonic resonators granting active polarization control via deterministic light-matter coupling.  
Our work is a first step towards highly scalable cavity quantum electrodynamics with engineered quantum emitters in two dimensional materials. 

\textbf{Funding:} State of Bavaria, H2020 European Research Council (ERC). National Research Foundation of Korea, Korean Government Grant NRF-2016R1C1B2007007.

\bibliography{library}
\end{document}